\documentclass[twocolumn,preprintnumbers]{revtex4-1}
\usepackage{amssymb, amsthm}
\usepackage{amsmath}    
\usepackage{mathrsfs}   
\usepackage{graphicx}   
\usepackage{color}      
\usepackage{footnote}   

\usepackage{slashed}    
\usepackage{subfigure}  
\usepackage{hyperref}   

\usepackage{rotating}   
\usepackage{multirow}   

\begin{document}
 
\preprint{NT@UW-16-06}
\title{Electrophobic Scalar Boson and Muonic Puzzles}
\author{Yu-Sheng Liu}\email{mestelqure@gmail.com} 
\author{David McKeen}\email{dmckeen@uw.edu}

\author{Gerald A. Miller}\email{miller@phys.washington.edu}
\affiliation{Department of Physics,	University of Washington, Seattle, Washington 98195-1560, U.S.A.}
\date{\today}

\begin{abstract}
A new scalar boson which couples to the muon and proton can simultaneously solve the proton radius puzzle and the muon anomalous magnetic moment discrepancy. Using a variety of measurements, we constrain the mass of this scalar and its couplings to the electron, muon, neutron, and proton. Making no assumptions about the underlying model, these constraints and the requirement that it solve both problems limit the mass of the scalar to between about 100~keV and 100~MeV. We identify two unexplored regions in the coupling constant-mass  plane. Potential future experiments and their implications for theories with mass-weighted lepton couplings are discussed.
\end{abstract}
\maketitle

 Recent measurements of the proton charge radius using the Lamb shift in muonic hydrogen are  troublingly discrepant with values extracted from 
 hydrogen spectroscopy and electron-proton scattering. Presently, the value from muonic hydrogen is 0.84087(39)~fm~\cite{Pohl:2010zza,Antognini:1900ns} while the CODATA average of data from hydrogen spectroscopy and $e$-$p$ scattering yields 0.8751(61)~fm~\cite{Mohr:2015ccw}; these differ at more than $5\sigma$. Although the discrepancy may arise from
  subtle lepton-nucleon non-perturbative effects within the standard model  or experimental uncertainties~\cite{Pohl:2013yb,Carlson:2015jba},  it could also be a signal of new physics 
  involving
  a violation of lepton universality. 

The muon anomalous magnetic moment provides  another potential signal of new physics. The BNL~\cite{Blum:2013xva}  measurement  differs from the standard model prediction by at least three standard deviations, $\Delta a_\mu=a_\mu^{\rm exp}-a_\mu^{\rm th}=287(80)\times 10^{-11}$~\cite{Davier:2010nc,Hagiwara:2011af}.

A new scalar boson,    
 $\phi$, that couples to the muon and proton could explain both the proton radius and $(g-2)_\mu$ puzzles~\cite{TuckerSmith:2010ra}. We investigate the couplings of this boson to standard model fermions, $f$, which appear as terms in the Lagrangian, $\mathcal{L}\supset e \epsilon_f\phi\bar ff$, where $\epsilon_f=g_f/e$ and $e$ is the electric charge of the proton.  Other authors have pursued this  basic idea, but  
  made further assumptions relating the couplings to different species; e.g. in~\cite{TuckerSmith:2010ra}, $\epsilon_p$ is taken equal to $\epsilon_\mu$ and in \cite{Izaguirre:2014cza}, mass-weighted couplings are assumed. References~\cite{TuckerSmith:2010ra} and~\cite{Izaguirre:2014cza} both neglect $\epsilon_n$. 
     We make no {\it a priori} assumptions regarding signs or magnitudes of the coupling constants. The Lamb shift in muonic hydrogen fixes $\epsilon_\mu$ and $\epsilon_p$ to have the same sign which, 
    we take to be positive. $\epsilon_e$ and $\epsilon_n$ are allowed to have either sign.

We focus on the scalar boson possibility because scalar exchange produces no hyperfine interaction, in accord with observation~\cite{Pohl:2010zza,Antognini:1900ns}.
 The emission of possible 
new vector particles becomes copious at high energies, and in the absence of an ultraviolet completion, is ruled out~\cite{Karshenboim:2014tka}.

\begin{figure}[t]
\centering
\includegraphics[scale=0.8]{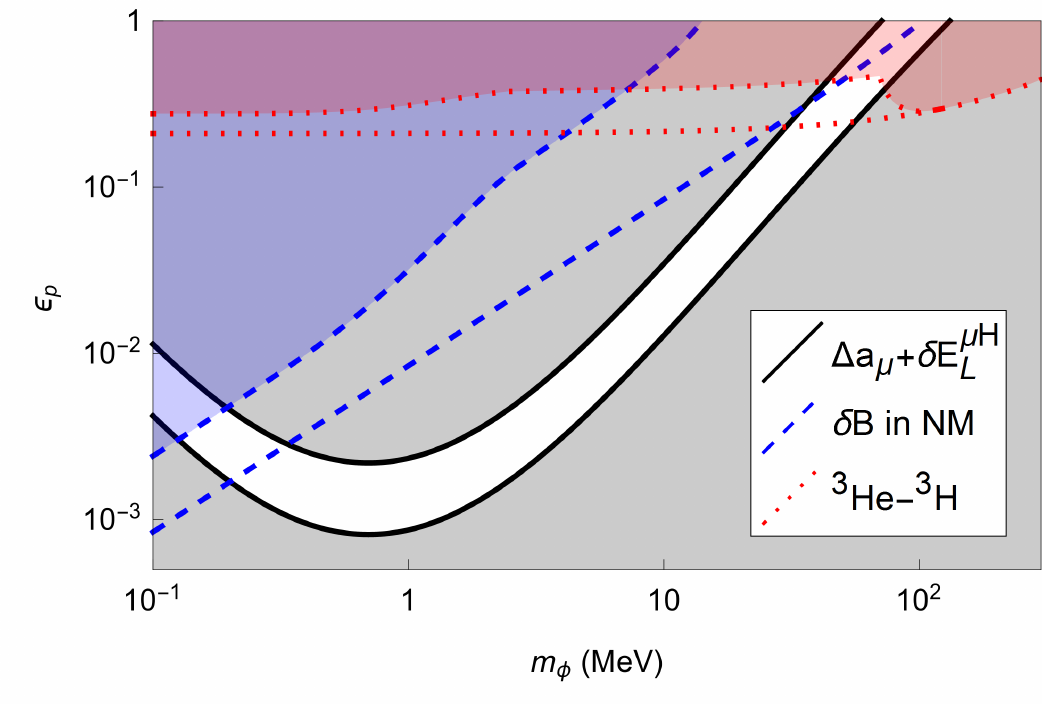}
\caption{\label{fig:gp_exclusion} (Color online) Exclusion (shaded regions) plot for $\epsilon_p$. The  region between the black lines is  allowed via 
  Eqs.~\ref{eq:g-2}-\ref{eq:deltadeltae}.  The dashed blue and dotted red lines represent  the constraints from nucleon binding energy in infinite nuclear matter and the $^3$He$-^3$H binding energy difference; isolated lines are derived using $\epsilon_n=0$ and the shaded regions are excluded using  the constraint on $\epsilon_n/\epsilon_p$ in Fig. \ref{fig:Rnp_exclusion}.} 
\end{figure}

Scalar boson exchange can account for  both  the proton radius puzzle and the $(g-2)_\mu$ discrepancy~\cite{TuckerSmith:2010ra}. The shift of the lepton $(\ell=\mu,e)$ muon's magnetic moment due to one-loop $\phi$ exchange is  { given by}~\cite{Jackiw:1972jz}
\begin{align}\label{eq:g-2}
\Delta a_\ell=\frac{\alpha \epsilon_\ell^2}{2\pi}\int_0^1dz\frac{(1-z)^{2}(1+z)}{(1-z)^{2}+(m_\phi/m_\ell)^{2}z}.
\end{align} 
Scalar exchange between fermions $f_1$ and $f_2$ leads to a Yukawa potential, $V(r)=-\epsilon_{f_1}\epsilon_{f_2}\alpha e^{-m_\phi r}/r$. In atomic systems, this leads to an additional contribution to the Lamb shift in the 2S-2P transition. For an (electronic or muonic) atom,  of  $A$ and $Z$ this shift is given by~\cite{Miller:2015yga} 
\begin{align}\label{eq:Lamb shift}
\delta E_L^{\ell\rm N}=-\frac{\alpha}{2a_{\ell\rm N}}\epsilon_\ell[Z\epsilon_p+(A-Z)\epsilon_n] f(a_{\ell\rm N}m_\phi) 
\end{align}
where $f(x)=x^2/(1+x)^4$~\cite{Barger:2010aj,TuckerSmith:2010ra}, with $a_{\ell\rm N}=(Z\alpha\,m_{\ell\rm N})^{-1}$ the Bohr radius and $m_{\ell\rm N}$ is the reduced mass of the lepton-nucleus system.
Throughout this paper we set
\begin{align}\label{eq:deltadeltae}
\Delta a_\mu=287(80)\times 10^{-11},\, \delta E_L^{\mu\rm H}=-0.307(56)~\rm meV
\end{align}
within two standard deviations. This value of  $\delta E_L^{\mu\rm H}$,  is the same as the energy shift caused by using the different values of the proton radius~\cite{Pohl:2010zza,Antognini:1900ns,Mohr:2015ccw,Antognini:2015moa} to explain the two discrepancies. This allows us to determine   both $\epsilon_p$ and $\epsilon_\mu$ as functions of $m_\phi$. The unshaded regions in Figs.~\ref{fig:gp_exclusion} and~\ref{fig:gmu_exclusion} show the values of $\epsilon_p$ and $\epsilon_\mu$, 
 as functions of the scalar's mass, that lead to the values of $\Delta a_\mu$ and $\delta E_L^{\mu\rm H}$ in Eq.~(\ref{eq:deltadeltae}).

We  study several observables sensitive to the couplings of the scalar to neutrons, $\epsilon_n$, and protons, $\epsilon_p$   to  obtain new 
bounds on $m_\phi$. 

\begin{figure}[t]
\centering
\includegraphics[scale=0.8]{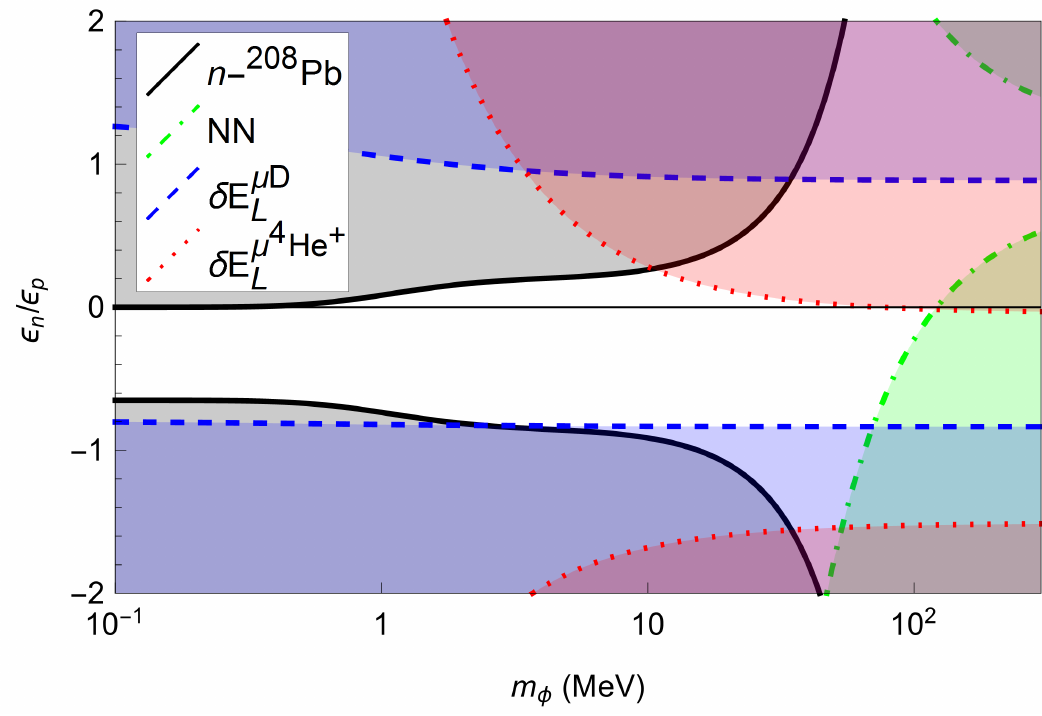}
\caption{\label{fig:Rnp_exclusion} (Color online) Exclusion (shaded regions) plot for $\epsilon_n/\epsilon_p$. The black, dashed blue, dotted red, and dotted dashed green lines correspond to the constraints from $n-^{208}$Pb scattering, $\mu$D Lamb shift, $\mu^4$He${^+}$ Lamb shift, and NN scattering length difference.}
\end{figure}
\begin{figure}[t]
\centering
\includegraphics[scale=0.8]{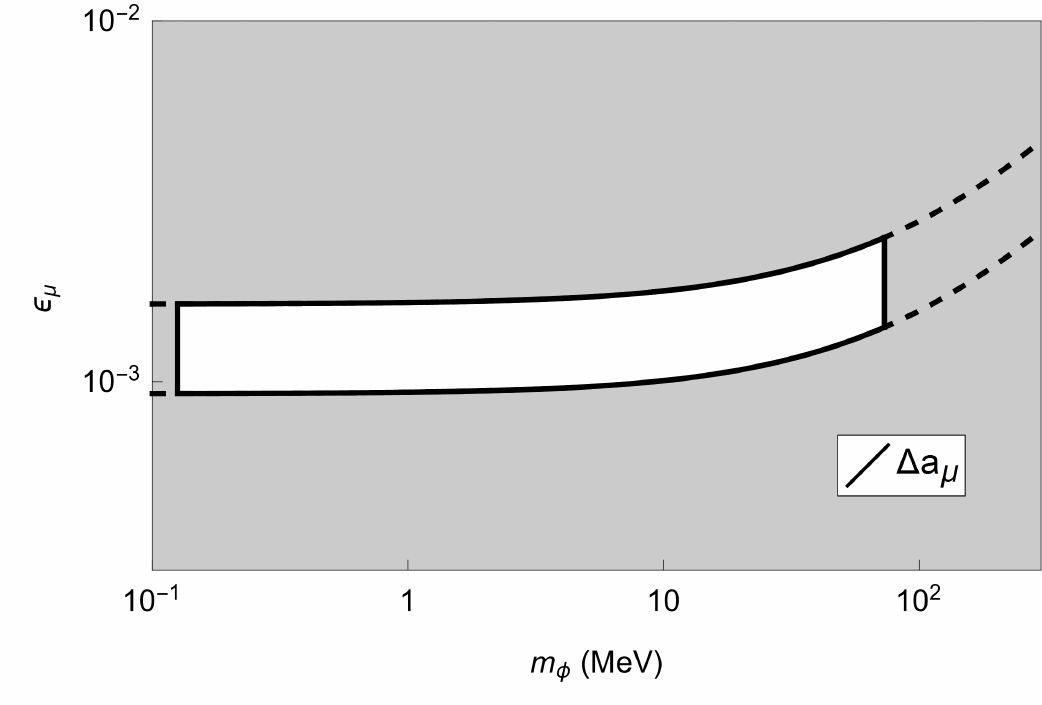}
\caption{\label{fig:gmu_exclusion} Exclusion (shaded region) plot for $\epsilon_\mu$. The region between the solid and dashed lines are obtained using $(g-2)_\mu$ Eq.~(\ref{eq:g-2}) with 2 S.D.. The restrictions on the values of $m_\phi$ in Fig.~\ref{fig:gp_exclusion} cause the region between the dashed lines to be excluded.}
\end{figure}

\begin{itemize}

\item {Low energy scattering of neutrons  on $^{208}$Pb}  has been used to constrain light force carriers coupled to nucleons~\cite{Leeb:1992qf} assuming a 
 coupling of a scalar to nucleons of $g_N$. Using  the replacement
\begin{align}\label{eq:g_neucleon replacement}
\frac{g_N^2}{e^2}\rightarrow \frac{A-Z}{A}\epsilon_n^2+\frac{Z}{A}\epsilon_p\epsilon_n
\end{align}
for scattering on a nucleus with atomic mass $A$ and atomic number $Z$, we 
separately constrain the coupling of a scalar to protons and neutrons.

\item The known NN charge-independence breaking scattering length difference, defined as $\Delta a=\bar{a}-a_{np}$, with $\bar{a}\equiv (a_{pp}+a_{nn})/2 $. The measured value $\Delta a^{\rm exp}=5.64(60)$ fm \cite{Machleidt:2001rw} is reproduced by a variety of known effects: $\Delta a^{\rm th}=5.6(5)$ fm \cite{Ericson:1983vw}. The existence of the scalar boson gives an additional  contribution 
\begin{align}
\Delta a_{\phi}=\bar{a}a_{np}M\int_0^\infty\Delta V\bar{u}u_{np}dr,
\end{align}
where M is the average of the nucleon mass; $\Delta V=-\frac{1}{2}\alpha(\epsilon_p-\epsilon_n)^2 e^{-m_\phi r}/r$; $u(r)$ is  the zero energy $^1S_0$ wave function, normalized so that $u(r)\rightarrow(1-r/a)$ as $r\rightarrow\infty$.  To avoid spoiling the agreement with experiment $\Delta a_{\phi}$ cannot be greater than 1.6 fm (using  2 S.D. as allowable).

\item  {The volume term in the semi-empirical mass formula gives  the binding energy per nucleon in $N=Z$ infinite  nuclear matter}. 
 Scalar boson exchange provides  an additional contribution. Using the Hartree approximation, which is accurate
if $m_\phi<100$ MeV) \cite{Mattuck:1976xt,SanjayReddy}, we find 
the average change in nucleon binding energy in infinite nuclear matter to be 
$({\delta B_p+\delta B_n})/{2}= {(g_p+g_n)^2\rho}/{4m_\phi^2}$  
which (with $\rho\approx0.08{\rm\;fm^{-3}}$) must not exceed 1 MeV to avoid problems with existing understanding of nuclear physics.

\item The difference in the binding energies of $^3$He and $^3$H of 763.76~keV is nicely explained by the effect of Coulomb interaction (693 keV) and charge asymmetry of nuclear forces  (about 68 keV)~\cite{Friar:1969zz,Friar:1978mr,Coon:1987kt,Miller:1990iz,Wiringa:2013fia}.  The contribution to the binding energy difference from the scalar boson can be estimated by using 
the  nuclear wave function extracted from elastic electron-nuclei scattering \cite{Friar:1978mr,Sick:2001rh,Juster:1985sd,Mccarthy:1977vd}.
We set constraints by  requiring  that this contribution not exceed 30~keV to maintain  the agreement between theory  and experiment.

\item 
We 
use  the preliminary results on the Lamb shifts in muonic deuterium and muonic $^4$He. For  $\mu$D a discrepancy similar to that of $\mu$H between the charge radius extracted via the Lamb shift of $\mu$D, $r^\mu_{\rm D}=2.1272(12)~\rm fm$~\cite{RandolfPohl} and the CODATA average from electronic measurements, $r_{\rm D}=2.1213(25)~\rm fm$~\cite{Mohr:2015ccw}, exists. This could be also be explained by a scalar coupled to muons that results in a change to the Lamb shift of $\delta E_L^{\mu\rm D}=-0.368(78)~\rm meV$~\cite{Antognini:2015moa,Krauth:2015nja}. The similarity of this shift to the one required in $\mu$H constrains the coupling of $\phi$ to the neutron. For $\mu^4$He, the radii extracted from the muonic Lamb shift measurement, $r^\mu_{^4\rm He}=1.677(1)~\rm fm$~\cite{AldoAntognini}, and elastic electron scattering, $r_{^4\rm He}=1.681(4)~\rm fm$~\cite{Sick:2014yha}, require the change in the Lamb shift due to $\phi$ exchange to be compatible with zero, $\delta E_L^{\mu^4{\rm He}^+}=-1.4(1.5)~\rm meV$~\cite{Antognini:2015moa}. Since these results are preliminary, we draw constraints at the $3\sigma$ level.  Using the ratio of  nuclear to hydrogen  Lamb  shifts  for D and He via Eq.~(\ref{eq:Lamb shift}) allows us to obtain 
  $\epsilon_n/\epsilon_p$ independently  of the value of $\epsilon_\mu$ and $\epsilon_p.$  
  We expect that publication of the D and $^4$He data would provide constraints at the $2\sigma$ level, thereby  narrowing  the allowed region by a factor of about 2/3  and changing   details of the borders of the allowed regions.

\end{itemize}

Using these observables (as  constrained by   Eqs.~(1-3))  we  limit the ratio of the coupling of $\phi$ to neutrons and protons, $\epsilon_n/\epsilon_p$, as shown in Fig.~\ref{fig:Rnp_exclusion}. If the couplings to neutron and proton are of the same sign, these constraints are  quite strong, driven by the { neutron-$^{208}$Pb}  scattering limits for $m_\phi\lesssim 10~\rm MeV$ and   the $\mu^4$He measurement for larger masses. If the couplings are of opposite sign, they   interfere destructively, masking the effects of the  $\phi$ and substantially weakening the limits on the magnitudes of $\epsilon_n,\epsilon_p$.

For a given value of $\epsilon_n/\epsilon_p$, we 
 use the shift of the binding energy in $N=Z$ nuclear matter and the difference in binding energies of $^3$H and $^3$He to constrain $\epsilon_p$. We show these bounds in Fig.~\ref{fig:gp_exclusion}, varying $\epsilon_n/\epsilon_p$ over its allowed range as a function of $m_\phi$. These measurements limit the mass of the scalar  that simultaneously explains  the proton radius and $(g-2)_\mu$ discrepancies to $100~{\rm keV}\lesssim m_\phi\lesssim 100~{\rm MeV}$. These upper and lower limits on the allowed value of $m_\phi$ are also indicated on the plot of the required values of $\epsilon_\mu$ in Fig.~\ref{fig:gmu_exclusion}.

\begin{figure}[t]
\centering
\includegraphics[scale=0.8]{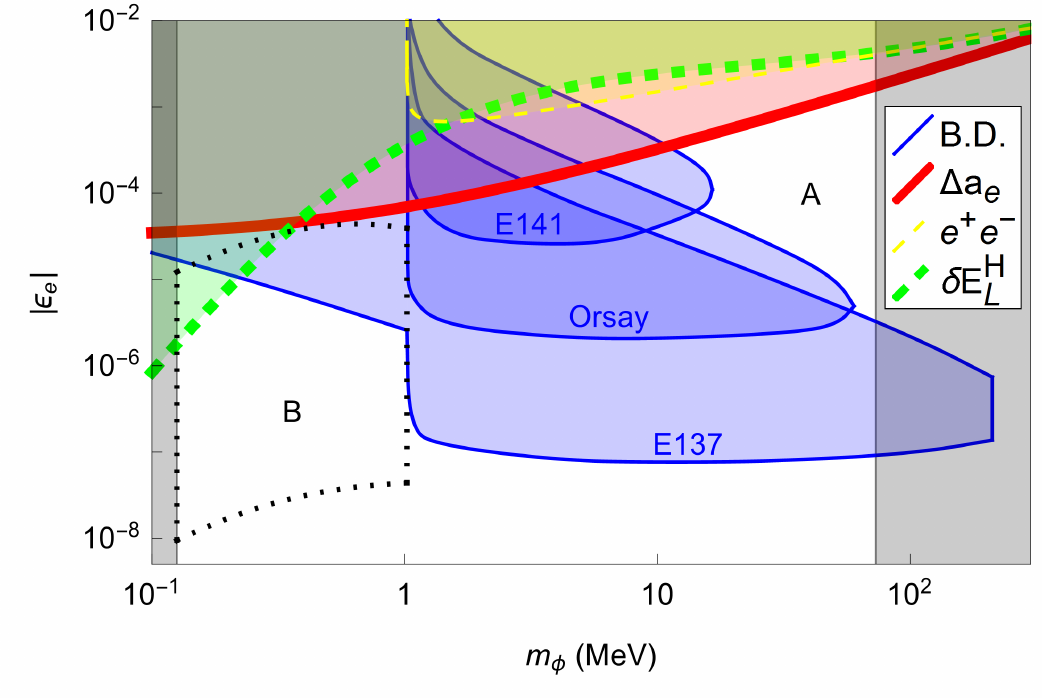}
\caption{\label{fig:ge_exclusion} (Color online) Exclusion (shaded regions) plot for $\epsilon_e$. The thick red, thin blue, thin dashed yellow, and thick dashed green lines correspond to the constraints from electron anomalous magnetic moment $(g-2)_e$, beam dump experiments, Bhabha scattering, and the Lamb shift of hydrogen. The region between the two vertical gray regions  are allowed using the scalar mass range from Fig. \ref{fig:gp_exclusion}.  Regions A and B 
could be covered by the  proposed experiments in: \cite{Bjorken:2009mm}, \cite{Izaguirre:2014cza} and  the study \cite{Essig:2013lka}.}
\end{figure}

We now explore the coupling of the scalar to electrons, which is of particular experimental importance because electrons are readily produced and comparatively simple to understand. The limits on the coupling $\epsilon_e$ are similar to many that have been placed on the dark photon in recent years (see, e.g.~\cite{Essig:2013lka}). Below, we describe the experimental quantities used to derive limits on the electron-scalar coupling.

Scalar exchange shifts the anomalous magnetic moment of the electron; see Eq.~(\ref{eq:g-2}). 
As emphasized in Ref.~\cite{Pospelov:2008zw}, the measurement of $(g-2)_e$ is currently used to extract the fine structure constant. A constraint on $\epsilon_e$ can therefore be derived by comparing the inferred value of $\alpha$ with a  value obtained from a measurement that isn't sensitive to the contribution of the scalar boson. We use the  precision study of $^{87}$Rb~\cite{Bouchendira:2010es}. Requiring that these two measurements agree implies that $\Delta a_e<1.5 \times 10^{-12}$ (2 S.D.).  

Bhabha scattering, $e^+e^-\rightarrow e^+e^-$, can be used to search for the scalar boson by looking for a resonance due to $\phi$ exchange. Motivated by earlier results from heavy-ion collisions near the Coulomb barrier, a GSI group
~\cite{Tsertos:1989gv} used a clean time-stable monoenergetic positron beam incident on a metallic Be foil. No resonances were observed at the 97\% C.L. within the experimental sensitivity of 0.5 b eV/sr (c.m.) for the energy-integrated differential cross section. Given the small value of $\epsilon_e$ the only relevant process is the $s$-channel exchange of a $\phi$ boson, which was not observed.  The experiment limits $|\epsilon_e|$  as shown in Fig.~4. 

Beam dump experiments have long been used to search for light, weakly coupled particles that decay to leptons or photons~\cite{Essig:2013lka,Bjorken:1988as,Bjorken:2009mm}. If coupled to electrons, $\phi$ bosons could be produced in such experiments and decay to $e^+e^-$ or $\gamma\gamma$ pairs depending on its mass. The production cross section for the scalar boson, not in the current literature, is discussed in a longer paper \cite{ysl16} to be presented  later.   Previous work~\cite{Bjorken:2009mm} simplified the evaluation of this cross section  by using the Weizsacker-Williams (WW) approximation,  by making further approximations to the phase space integral, assuming  that the mass of the new particle is much greater than electron mass,  and can't be used if $m_\phi<2m_e$.  Our numerical evaluations~\cite{ysl16}  do not  use these assumptions and thereby allow us to cover the entire mass range shown in Fig.~4. In particular, we find that  the approximations of~\cite{Bjorken:2009mm}  have significant   errors for $m_\phi>10 $ MeV.  Our analysis uses data from  the electron beam dump experiments E137~\cite{Bjorken:1988as}, E141~\cite{Riordan:1987aw}, and Orsay~\cite{Davier:1989wz}.

In addition to muonic atoms, scalar exchange will affect the Lamb shift in ordinary electronic atoms. To set limits on the coupling, following~\cite{Miller:2011yw,Miller:2012ht,Miller:2012ne}, we require that the change to the Lamb shift in hydrogen is $\delta E_L^{\rm H}<14~\rm kHz$~\cite{Eides:2000xc}(2 S.D.). 

In Fig.~\ref{fig:ge_exclusion}, we present the constraints on the coupling to electrons, $\epsilon_e$, as a function of $m_\phi$ from these observables. In addition we indicate (via two dashed vertical lines)  the allowed mass range for $\phi$, taken from Fig.~\ref{fig:gp_exclusion}.

We label two allowed regions in the $(m_\phi,\epsilon_e)$ plane in Fig.~\ref{fig:ge_exclusion}: A, where $10~{\rm MeV}\lesssim m_\phi\lesssim 70~{\rm MeV}$, $10^{-6}\lesssim\epsilon_e\lesssim10^{-3}$, and B, where $100~{\rm keV}\lesssim m_\phi\lesssim 1~{\rm MeV}$, $10^{-8}\lesssim\epsilon_e\lesssim10^{-5}$. There are a number of planned electron scattering experiments that will be sensitive to scalars with parameters in Region A, such as, e.g., APEX~\cite{Essig:2010xa,*Abrahamyan:2011gv}, HPS~\cite{Battaglieri:2014hga}, DarkLight~\cite{Freytsis:2009bh,*Cowan:2013mma}, VEPP-3~\cite{Wojtsekhowski:2012zq}, and MAMI or MESA~\cite{Beranek:2013yqa}. As studied in Ref.~\cite{Izaguirre:2014cza}, region B can be probed by looking for scalars produced in the nuclear de-excitation of an excited state of $^{16}$O. We have translated this region of couplings $10^{-11}\leq\epsilon_p\epsilon_e\leq10^{-7}$ from Ref.~\cite{Izaguirre:2014cza} to show on our plot by taking $\epsilon_p\rightarrow\epsilon_p+\epsilon_n$, 
using $\epsilon_n/\epsilon_p$ from Fig.~2 and fixing $\epsilon_p$ according to (\ref{eq:deltadeltae}).

We do not show limits derived from stellar cooling that are sensitive to $m_\phi\lesssim 200~\rm keV$~\cite{An:2013yfc} since the lower bound on the mass is similar to the one we have derived. Additionally, we note that constraints from cooling of supernovae do not appear in Fig.~\ref{fig:ge_exclusion}. This is because the required value of $g_p$ is always large enough to keep any scalars produced trapped in supernova, rendering cooling considerations moot~\cite{Rrapaj:2015wgs}. The effects of this scalar boson may have cosmological consequences,  beyond the scope of this article.  

We summarize the parameter space (see  also Table I): 
\begin{enumerate}
\item The range of allowed $m_\phi$ is widened from a narrow region around 1~MeV in~\cite{TuckerSmith:2010ra} to  the region from about 130~keV to 73 MeV  by allowing $\epsilon_p\ne\epsilon_\mu$ .

\item We carefully deal with $\epsilon_n$ instead of neglecting it. In particular, as seen in  Fig.~\ref{fig:gp_exclusion}, allowing $\epsilon_n$ to be of the opposite sign of $\epsilon_p$ opens up the parameter space.
\item The constraint on $\epsilon_e$ at $m_\phi=1$ MeV is improved by two orders of magnitude compared with \cite{TuckerSmith:2010ra} by using electron beam dump experiments.
\item Near the maximum allowed $m_\phi\sim 70$ MeV, the allowed couplings are relatively large, $|\epsilon_e|<1.8\times 10^{-3}$; $10^{-3}<\epsilon_\mu<2\times 10^{-3}$; $\epsilon_p\lesssim 0.4$; $-0.3\lesssim\epsilon_p\lesssim 0$,  providing ample opportunity to test this solution.
\end{enumerate}

\begin{table*}[t]
\centering
\caption{\label{table:couplings} Allowed coupling with various scalar mass: numbers in the parentheses are 1 S.D..}
\begin{tabular}{|c|c|c|c|c|}
\hline $m_\phi$ (MeV) & $|\epsilon_e|$ & $\epsilon_\mu$ & $\epsilon_p$ & $\epsilon_n$\\
\hline 0.13 & $<2.0\times 10^{-6}$ & $1.29(18)\times 10^{-3}$ & $3.0\times 10^{-3}$ 		& $-2.0\times 10^{-3}$ to $2.8\times 10^{-7}$\\
\hline 1 	& $<2.6\times 10^{-6}$ & $1.30(18)\times 10^{-3}$ & $1.60(37)\times 10^{-3}$ 	& $-1.7\times 10^{-3}$ to $2.0\times 10^{-4}$\\
\hline 10	& $<7.6\times 10^{-8}$ & $1.40(20)\times 10^{-3}$ & $2.37(54)\times 10^{-2}$ 	& $-2.9\times 10^{-2}$ to $9.1\times 10^{-3}$\\
\hline 73   & \begin{tabular}{c}$<9.1\times 10^{-8}$\\ $3.3\times 10^{-6}$ to $1.8\times 10^{-3}$\end{tabular} 
									& $1.96(27)\times 10^{-3}$ & 0.39 						& $-0.29$ to $5.6\times 10^{-4}$\\
\hline
\end{tabular}
\end{table*}

Our discussion thus far has been purely phenomenological, with no particular  UV completion in mind to relate the couplings of fermions with the same quantum numbers (here  the electron and muon). From the model-building point of view, there are motivations that the couplings of $\phi$ to fermions in the same family are mass-weighted--in particular, for the leptons, $|\epsilon_\mu/\epsilon_e|=(m_\mu/m_e)^n$ with $n\ge 1$. This is because, generally, coupling fermions to new scalars below the electroweak scale leads to large flavor-changing neutral currents (FCNCs) that are very strongly constrained, e.g. in the lepton sector by null searches for $\mu\to e$ conversion, $\mu\to 3e$, or $\mu\to e\gamma$. A phenomenological ansatz for the structure of the $\phi$'s couplings to fermions that avoids this problem is that its Yukawa matrix is proportional to that of the Higgs. This scenario has been termed minimal flavor violation (MFV), see e.g.~\cite{Cirigliano:2005ck}. In that case, both the Higgs and $\phi$ couplings are simultaneously diagonalized and new FCNCs are absent. The main phenomenological consequence of this is that $\phi$'s coupling to a lepton is proportional to a power of that lepton's mass, $\epsilon_\ell\propto m_\ell^n$ with $n\ge 1$. 
 In the context of a given model, i.e. for fixed $n$, we can relate Figs.~\ref{fig:gmu_exclusion} and~\ref{fig:ge_exclusion}.  Region A largely corresponds to $0<n\lesssim 1$, which, 
is less well-motivated from a model building perspective. $1\lesssim n\lesssim 2$ is well-motivated and fits into Region B. To obtain $\epsilon_e\lesssim10^{-7}$, $n\gtrsim 2$ is required.  All of the  allowed values of $\epsilon_e$ are smaller than the required value of $\epsilon_\mu$, thus the name electrophobic scalar boson is applicable.

Building  a complete model, valid at high energy scales,  leading  to interactions at low energies  is not our purpose. However,  we    outline one  simple possibility. In the lepton sector, couplings to $\phi$ could arise 
 through mixing obtained via a lepton-specific two Higgs doublet model, which would automatically yield MFV~\cite{Batell:2016ove}.  
 In the quark sector, coupling to a light boson via mixing with a Higgs is very tightly constrained by null results in $K\to\pi$ and $B$ meson decays (see, e.g.,~Ref.~\cite{Batell:2009jf}) decays. However, as in Ref.~\cite{Fox:2011qd}, heavy vector-like quarks that couple to $\phi$ and mix primarily with right-handed quarks of the first generation due to a family symmetry are a possibility. The coupling strength of $\phi$ to $u$ and $d$ quarks could differ leading to different couplings to neutrons and protons. If, e.g., $g_d/g_u\sim-0.8$ then $g_n/g_p\sim-0.5$, which, as we see in Fig.~\ref{fig:Rnp_exclusion}, is comparatively less constrained.

The existence of a scalar boson that couples to muons and protons accounts for the proton radius puzzle and the present discrepancy in the muon anomalous magnetic moment.  Many previous experiments could have detected  this particle, but none did. Nevertheless, regions A and B in Fig.~\ref{fig:ge_exclusion} remain open for discovery.

For masses $m_\phi$ near the allowable maximum,  the value of $\epsilon_p$ can be as large as about  0.4. Such a  coupling  could be probed with   proton experiments, such as threshold $\phi$ production in $pp$ interactions. Proton or muon beam dump experiments could also be used~\cite{Gardner:2015wea}. Can one increase the accuracy 
of the neutron-nucleus experiments?
For experiments involving muons, one might think of using muon beam dump experiments, such as the COMPASS experiment as proposed in~\cite{Essig:2010gu}. The MUSE experiment~\cite{Gilman:2013eiv}  plans to measure  $\mu^\pm$ and $e^\pm$-p elastic scattering at low energies. Our hypothesis regarding the $\phi$ leads to a prediction for the MUSE experiment even though its direct effect on the scattering will be very  small~\cite{Liu:2015sba}: the MUSE experiment will observe  the same `large' value of the proton radius for all of the probes.   Another possibility is to study the spectroscopy of muonium (the bound state of $e^-$ and $\mu^+$) or true muonium (the bound state of $\mu^-$ and $\mu^+$). 
Perhaps 
the best way to test the existence of this particle would be an improved measurement of the muon anomalous magnetic moment \cite{Hertzog:2015jru}. The existence of a particle with such a limited role may  seem improbable, considering the present state of knowledge. However, such an existence is not ruled out.  

\section*{Acknowledgments.} We thank M. Barton-Rowledge, J. Detwiler, R. Machleidt, S. Reddy, and M. J. Savage for invaluable discussions and suggestions. The work of G. A. M. and Y.-S. L. was supported by the U. S. Department of Energy Office of Science, Office of Nuclear Physics under Award Number DE-FG02-97ER-41014. The work of D.~M. was supported by the U.S. Department of Energy under Grant No. DE-FG02-96ER-40956. 

\bibliography{scalar_letter}
 \end{document}